\documentclass{llncs}

\usepackage[utf8]{inputenc}
\usepackage[T1]{fontenc}
\usepackage[svgnames]{xcolor}
\usepackage{pifont}
\usepackage[abbreviations]{foreign}
\usepackage{acronym}
\usepackage{booktabs}
\usepackage[caption=false]{subfig}
\usepackage[binary-units,per-mode=symbol]{siunitx}
\usepackage{rotating}
\usepackage{floatrow}
\usepackage[hidelinks,bookmarksdepth=2]{hyperref}
\usepackage{csquotes}
\usepackage{tikz}

\newcommand{\SYS}{\textsc{Stress-SGX}\xspace}
\newcommand{\vanilla}{\textsc{Stress-ng}\xspace}

\acrodef{SGX}{Software Guard Extensions}
\acrodef{TEE}{Trusted Execution Environment}
\acrodef{SDK}{Software Development Kit}
\acrodefindefinite{SDK}{an}{a}
\acrodef{PSW}{Platform Software}
 
\hypersetup{
	pdftitle={\SYS: Load and stress your enclaves for fun and profit},
	pdfkeywords={Intel SGX, Load, Stress, Benchmark},
	pdfauthor={Sébastien Vaucher, Valerio Schiavoni, Pascal Felber},
}

\begin{document}

\mainmatter
\title{\SYS: Load and Stress your\\ Enclaves for Fun and Profit}

\author{Sébastien~Vaucher \and
Valerio~Schiavoni \and
Pascal~Felber}

\institute{University of Neuchâtel, Neuchâtel, Switzerland\\
\email{\{\href{mailto:sebastien.vaucher@unine.ch}{sebastien.vaucher},\href{mailto:valerio.schiavoni@unine.ch}{valerio.schiavoni},\href{mailto:pascal.felber@unine.ch}{pascal.felber}\}@unine.ch}}

\maketitle

\begin{abstract}%
\begin{tikzpicture}[remember picture, overlay]
\node[anchor=north,xshift=60pt,rotate=90,black!70] at (current page.west) {\parbox{420pt}{%
	This is a post-peer-review, pre-copyedit version of an article published in \enquote{Networked Systems}. The final authenticated version is available online at https://doi.org/10.1007/978-3-030-05529-5\_24.
}};
\end{tikzpicture}%
The latest generation of Intel processors supports \ac{SGX}, a set of instructions that implements \iac{TEE} right inside the CPU, by means of so-called enclaves.
This paper presents \SYS, an easy-to-use stress-test tool to evaluate the performance of SGX-enabled nodes.
We build on top of the popular \vanilla tool, while only keeping the workload injectors (\emph{stressors}) that are meaningful in the SGX context.
We report on several insights and lessons learned about porting legacy code to run inside an SGX enclave, as well as the limitations introduced by this process.
Finally, we use \SYS to conduct a study comparing the performance of different SGX-enabled machines.

\end{abstract}

\acresetall

\section{Introduction}
\label{sec:intro}

The latest generation of Intel processors (starting from the Skylake microarchitecture) features a new set of instructions: \ac{SGX}.
This instruction set allows programs to execute securely inside hardware \emph{enclaves}, hence creating \iac{TEE}.
Specifically, \ac{SGX} enclaves protect the code from external threats, including privileged system software.
Given the novelty of the technology (the first compatible CPU was released in August 2015) and the lack of in-depth literature, it is still challenging to validate the performance of code running inside enclaves.
It becomes even harder to study the problem under varying conditions, such as different hardware revisions or workloads. 
Moreover, as we show in this paper, microcode updates issued by CPU vendors can introduce performance degradations that are difficult to promptly detect.

The main contribution of this paper is \SYS, a stress tool capable of artificially putting SGX enclaves under high load.
It supports workloads of different nature, as explained in \autoref{sec:impl}.
As far as we know, \SYS is the first tool that can be leveraged to induce hybrid workloads (with and without SGX).
We believe that many researchers can benefit from using our tool, simplifying design plans for their evaluation settings.
\SYS is GPL-licensed free software\footnote{\url{https://github.com/sebva/stress-sgx}}.

\paragraph{\bf Motivating scenarios.}
We introduce three use-cases for \SYS and how researchers could leverage this tool.
First, there exist several real-world cluster traces readily available for research, such as the ones released by Google Borg~\cite{clusterdata:Wilkes2011} or Microsoft Azure~\cite{Cortez:2017:RCU:3132747.3132772}.
However, these traces do not have the same properties as typical SGX workloads.
Due to anonymization, hardware characteristics of the original machines remain undisclosed.
Finally, the nature of jobs deployed on these clusters is also confidential.
\SYS allows to directly map behaviors described in such traces onto similar, SGX-specific ones.

A second motivating scenario stems from the availability of a range of SGX-enabled CPUs on the market, each with their own characteristics and associated performance.
Apart from the expected differences due to hardware revision, cache size, available instruction sets or frequency, one still needs to hand-craft micro-benchmarks in order to evaluate SGX-specific performance.
\SYS facilitates this process by providing the same interface as \vanilla~\cite{stress-ng} to inject load on a CPU under controlled conditions. 

A third motivating scenario directly derives from the necessity to measure the electric energy consumption of code executing inside enclaves, by means of software~\cite{bourdon2013powerapi} or hardware power meters.
\SYS offers a common code-base for both SGX and native contexts, making it easier to isolate the energy requirements of SGX-specific workloads.

\paragraph{\bf Roadmap.}

The remainder of the paper is organized as follows: \autoref{sec:impl} describes some implementation details and lessons learned.
\autoref{sec:eval} presents our preliminary evaluation. 
Finally, \autoref{sec:conclusion} discusses future work and concludes.

\section{Implementation}
\label{sec:impl}

\SYS is implemented as a fork of \vanilla (version 0.09.10).
It directly reuses the same compilation and runtime foundations.
From a user perspective, SGX-enabled stressors (specific pieces of executable code that exercise a given CPU functionality or low-level operation) are selected by specifying command-line options starting with \texttt{-{}-sgx}.
We concentrate on porting CPU stressing methods to run in an enclave in a way that makes both native and SGX versions comparable from a performance standpoint.
All features offered by the CPU stressors shipped with \vanilla are supported by \SYS, with the exception of partial load.
Specifying a fixed load percentage is not possible (\ie it is locked to \SI{100}{\percent}), because this feature is based on precise timing, which is currently not available inside the enclave.
Our \SYS prototype supports \num{54} enclave-enabled stress methods out of the \num{68} currently shipped with \vanilla.
The exhaustive list is presented in \autoref{tab:stressors}.
We note that support for each stressor depends on the availability of its required functionalities within the enclave, as well as their support by the \ac{SDK}.

\newcommand{\y}{\color{ForestGreen}{\ding{51}}}
\newcommand{\n}{\color{FireBrick}{\ding{55}}}

\newcommand*\rot{\rotatebox{90}}
\renewcommand{\tabcolsep}{0.9pt}
\begin{table}[t]
	\centering
	\caption{List of stressors supported by \SYS. ``{\color{FireBrick}\ding{55}}'' indicates a stressor available in \vanilla that is not SGX-compatible. \texttt{f}=float, \texttt{d}=double, \texttt{ld}=longdouble.}
	\label{tab:stressors}
	\resizebox{\textwidth}{!}{%
	\begin{tabular}{lllllllllllllllllllllllllllllllllllllllllllllllllllllll}
	\toprule
	\rot{ackermann} &
	\rot{bitops} &
	\rot{callfunc} &
	\rot{complex[f,d,ld]} &
	\rot{correlate} &
	\rot{crc16} &
	\rot{decimal[32,64,128]} &
	\rot{dither} &
	\rot{djb2a} &
	\rot{double} &
	\rot{euler} &
	\rot{explog} &
	\rot{factorial} &
	\rot{fft} &
	\rot{fibonacci} &
	\rot{float} &
	\rot{fnv1a} &
	\rot{gamma} &
	\rot{gcd} &
	\rot{gray} &
	\rot{hamming} &
	\rot{hanoi} &
	\rot{hyperbolic} &
	\rot{idct} &
	\rot{int8} &
	\rot{int16} &
	\rot{int32[f,d,ld]} &
	\rot{int64[f,d,ld]} &
	\rot{int128[f,d,ld]} &
	\rot{jenkin} &
	\rot{jmp} &
	\rot{ln2} &
	\rot{longdouble} &
	\rot{loop} &
	\rot{matrixprod} &
	\rot{nsqrt} &
	\rot{ocall} &
	\rot{omega} &
	\rot{parity} &
	\rot{phi} &
	\rot{pi} &
	\rot{pjw} &
	\rot{prime} &
	\rot{psi} &
	\rot{queens} &
	\rot{rand} &
	\rot{rand48} &
	\rot{rgb} &
	\rot{sdbm} &
	\rot{sieve} &
	\rot{stats} &
	\rot{sqrt} &
	\rot{trig} &
	\rot{union} &
	\rot{zeta} \\

	\midrule
	\y & \y & \y & \n & \y & \y & \n & \y& \y&\y &\y &\y & \y & \y & \y & \y &\y &\y &\y &\y &\y & \y&\y &\y &\y &\y &\y &\y &\n & \y&\y &\y &\y &\y &\n & \y & \y& \y&\y &\y &\y &\y &\y &\y &\y &\y & \n &\y &\y & \n & \y &\y &\y & \y & \y \\
	\bottomrule
	\end{tabular}}
\end{table}
 
\paragraph{\bf Porting CPU stressors to run inside an enclave.}

The Intel \ac{SGX} \ac{SDK} is designed in a way that allows existing code to be ported to run in an enclave with reasonable engineering efforts~\cite{sgx-sdk}.
We take advantage of this fact to port the CPU stressors of \vanilla in our own \SYS fork.
We detail here the particular complications that we encountered throughout this porting effort, and the specific adjustments applied to solve each of them.

After creating an enclave using the template given by the \ac{SDK}, we copy the relevant source code for the CPU stressors in the aforementioned template.
\vanilla defines several macros in its global header file.
It is not possible to include the code \emph{verbatim} as it depends on numerous system-specific features.
Our solution is to define the needed symbols on a case-by-case basis.

The next obstacle is the need to split the code in \emph{trusted} and \emph{untrusted} parts.
We decompose the code in a way that limits the number of enclave transitions (\ie entering and exiting the enclave) required to run a stressor.
Transitions are costly~\cite{Orenbach:2017:EEO:3064176.3064219}, so it is crucial that none happen while stressing is in progress, to ensure a consistent behavior.

The user can gracefully abort the execution of \SYS using standard Linux signals such as \texttt{SIGINT}.
\vanilla includes a mechanism to catch the majority of signals and react accordingly for its built-in stressors.
We leverage the fact that SGX enclaves can access the \emph{untrusted} memory of their enclosing process to pass a pointer to the \texttt{g\_keep\_stressing\_flag} variable to the enclave.
This variable is later used to indicate when to stop the execution of stressors.
Code running inside the enclave periodically polls the flag, and stops the execution if asked by the user.
The same flag is also used to make a stressor run for a given duration.
Timekeeping is done outside the enclave, with the indication to stop the execution notified by changing the value of the variable.

\paragraph{\bf Ensuring byte-per-byte equivalence of native and SGX code.}

During the initial testing phase of \SYS, we observed vast differences in performance for the same stressor executed in native and enclave modes.
As a matter of fact, while the source code was identical in both instances, the resulting compiled binaries slightly differ.
We believe that these slight differences are inevitable, as different linking rules are needed by each execution mode.

Conveniently, an enclave compiled using the official \ac{SGX} \ac{SDK} will produce a statically-compiled shared library.
We leverage this aspect to guarantee that both native and SGX versions of a stressor execute a perfectly identical binary by dynamically linking this shared object.
Choosing this optional approach limits \SYS to stressors that are available in enclave mode.

\begin{figure}[t]\TopFloatBoxes
\begin{floatrow}
	\ttabbox{%
	\resizebox{209.1pt}{!}{%
	\begin{tabular}{lllcSS}
		\toprule
		&&&& \multicolumn{2}{c}{Freq. [\si{\giga\hertz}]} \\
		\cmidrule{5-6}
		Category & Model & Processor & {Cores} & {Base} & {Max.} \\
		\midrule
		Server & Supermicro 5019S-M2 & Xeon E3-1275\,v6 & 4 & 3.8 & 4.2 \\
		Desktop & Dell Optiplex 7040 & Core i7-6700 & 4 & 3.4 & 4.0 \\
		NUC & Intel NUC7i7BNHX1 & Core i7-7567U & 2 & 3.5 & 4.0 \\
		Stick & Intel STK2m3W64CC & Core m3-6Y30 & 2 & 0.9 & 2.2 \\
		\bottomrule
	\end{tabular}}%
	}{%
		\caption{Hardware characteristics of our test machines. All processors are made by Intel.}%
		\label{tab:machines}%
	}
	\ffigbox[\FBwidth]{%
		\raggedleft%
		\includegraphics{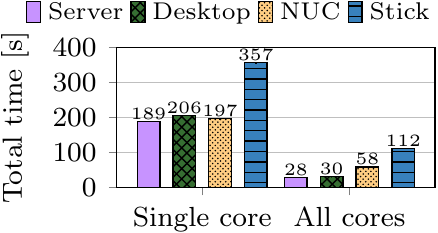}%
	}{%
		\caption{Time needed to perform \num{100000000} enclave transitions.}%
		\label{fig:ocall}%
	}
\end{floatrow}
\end{figure}

\section{Evaluation}
\label{sec:eval}
\hyphenation{pre-sents}

This section presents our preliminary evaluation of \SYS.
\autoref{tab:machines} lists the SGX machines used for our experiments.
We chose these machines due to their different form factors, hardware features, and widespread market availability. 
We expect them to represent a meaningful sample of SGX machines that are in use nowadays.   
These machines are configured to run Ubuntu Linux 17.10, along with v2.0 of the Intel SGX \ac{SDK}, \ac{PSW} and driver.

\paragraph{\bf Cost of enclave transitions.}
When programming for SGX, it is important to keep in mind that the cost required to enter and exit an enclave is significant~\cite{Orenbach:2017:EEO:3064176.3064219}.
\autoref{fig:ocall} presents the time needed to perform \num{100} million enclave transitions on a single core at a time (left) and on all available cores (right).
As expected, the server-class machine is consistently faster than the other ones, while the Intel Compute Stick performs the worst. 
We also observe marginal differences in single-core performance, in which case the processor can run at its maximal turbo frequency.
Given the notable price difference between the various machines, the Intel NUC offers the best price-performance ratio.

\begin{figure}[t]
	\centering
	\includegraphics[width=\textwidth]{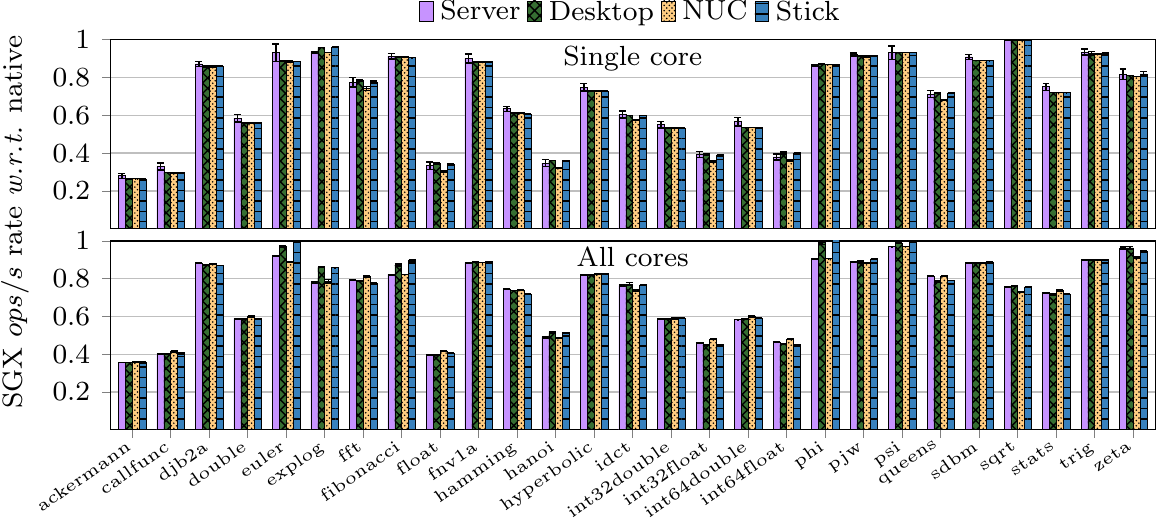}
	\caption{SGX performance compared to native on different types of computers, using January 2018 microcode (previous microcode had performance similar to native).}
	\label{fig:performance}
\end{figure}

\paragraph{\bf Cost of latest microcode update.} The second suite of microbenchmarks highlights a rather surprising effect of a recent microcode update, recently issued by Intel (on 2018-01-08) to mitigate the Spectre attack~\cite{kocher2018spectre}.
To observe the performance impact of this microcode update, we execute all \num{54} supported stressors before and after the microcode update.
Using the previous microcode, all stressors display the same performance in SGX and native modes.
Under the updated microcode, on the other hand, we observe a significant difference in SGX \emph{versus} native performance.
\autoref{fig:performance} presents the results for the 27 tests for which SGX performance is affected.
We measure slowdowns up to $\num{3.8}\times$ (\texttt{ackermann} running on a single core).
Given the undisclosed nature of microcode updates, it is difficult to identify the root cause of this performance degradation.

 \acresetall

\section{Conclusion and Future Work}
\label{sec:conclusion}

The expanding availability of SGX-enabled machines calls for new tools to evaluate the performance of \emph{secure clusters}.
The current lack of ready-to-deploy SGX applications usually forces researchers to implement single-use workloads.
This paper presented \SYS, an easy-to-use tool capable of stressing SGX enclaves and report on diverse metrics.
We plan to extend our prototype along the following directions: first, we intend to evaluate the performance of reading and writing encrypted memory using the memory stressors of \vanilla, ported in \SYS (not presented in this paper due to lack of space).
Second, we will integrate these stressors within a large-scale monitoring and performance framework for containerized microservices, to easily monitor performance regressions.

\paragraph{\bf Acknowledgement.} The research leading to these results has received funding from the European Union's Horizon 2020 research and innovation programme under the LEGaTO Project (\href{https://legato-project.eu/}{legato-project.eu}), grant agreement No~780681.

\bibliographystyle{splncs04}
\bibliography{references}

\end{document}